\documentclass[aps,prb,floats,twocolumn]{revtex4}
\usepackage{epsfig}
\renewcommand{\vec}{\mathbf}

\begin{document}
\title{On the origin of temperature dependence of interlayer exchange
  coupling in metallic trilayers} 
\author{S.~Schwieger and W.~Nolting}
\affiliation{Humboldt-Universit{\"a}t zu Berlin, Institut f{\"u}r Physik,
  Newtonstr. 15, 12489 Berlin}
\begin{abstract}
We study the influence of collective magnetic excitations on the
interlayer exchange coupling (IEC) in metallic multilayers. The results
are compared to other models that explain the temperature dependence of
the IEC by mechanisms within the spacer or at the interfaces of the
multilayers. As a main result we find that the reduction of the IEC with
temperature shows practically the same functional dependence in all
models. On the other hand the influence of the spacer thickness, the
magnetic material, and an external field are quite different. Based on
these considerations we propose experiments, that are able to determine the
dominating mechanism that reduces the IEC at finite temperatures.
\end{abstract}
\maketitle
\section{introduction}
A lot of aspects of the coupling of two magnetic layers separated by a
paramagnetic, metallic spacer are well
understood today\cite{Bru95}. The coupling is caused by spin dependent reflections
of spacer electrons at the interfaces. It oscillates with the spacer
thickness $D$. The periods are determined by the
spacer, namely by stationary Fermi surface spanning vectors in growth
direction. This are vectors parallel to the film normal that connect two
points on the Fermi surface and have a vanishing first derivative with
respect to the planar components of the Fermi vectors.\\
However, the origin of the temperature dependence is still under
discussion. Up to now it is not clear if the temperature dependence
is governed by effects within the spacer, at the interface or within the
magnetic layers. There are several proposals for mechanisms  reducing the
coupling at finite temperatures.
\begin{itemize}
\item[(i)] spacer contribution\\
One reason of the reduced IEC is the softening of the Fermi edge
at higher temperatures, which makes the coupling mechanism less
effective. This was proposed by Bruno and Chappert \cite{BrC91} and
Edwards et.al. \cite{EMM91} It leads to a certain temperature dependent
factor for each oscillation period.

\item[(ii)] interface contribution\\
The argument $\phi_\sigma$ of the complex reflection coefficients $r_\sigma=|r_\sigma|e^{i\phi_\sigma}$ at the
spacer/magnet interface may be highly energy dependent. This gives rise to an
additional temperature dependence of the IEC since the energy interval of
interest around the Fermi energy increases with
temperature\cite{AMV96,CAF99}. The same may in principle apply to the
norm of $r_\sigma$\cite{Bru99}.
A rather obvious effect is the reduction of the spin asymmetry of the reflection coefficient
$\Delta r=r_\uparrow-r_\downarrow$ with temperature. 
\item[(iii)] magnetic layers\\
Collective excitations within the magnetic layers reduce their free
energy. Since the layers are coupled the excitations depend on the
angle between the magnetization vectors of both layers. 
Thus the reduction of the free energy will be different for parallel and
antiparallel alignment of the magnetic layers.
This difference 
\begin{equation}
\Delta F_{\rm mag}(T)=F_{\rm mag}^{\uparrow\uparrow}(T)-F_{\rm
  mag}^{\uparrow\downarrow}(T)\nonumber
\end{equation}
contributes to the temperature dependence of the IEC.\\
\end{itemize}
The first two contributions are closely associated with the coupling
mechanism. The third effect works rather parallel to the
coupling mechanism itself, but nevertheless has consequences for the
amount of energy achieved by the coupling.\\
It is the aim of this paper to study the role of the different
contributions to the temperature dependence of the IEC. 
Thereto we have to gain explicit expressions for
case (iii). The first two contributions can be described in the
frame of ab initio theory combined with Fermi liquid theory\cite{KDT00,DKB99} as
well as in an quantum well picture\cite{Bru95}. They are thoroughly discussed
in literature. The third mechanism is due to collective magnetic
excitations which are beyond the scope of these theories. We derive the
expressions using a Heisenberg model which is best suited to describe
the low energy spin wave excitations within the magnetic layers.\\
The paper is organized as follows: In the next section we review and discuss the
spacer and the interface contribution. In section 3 we introduce
our model system, derive the expressions for the magnetic contribution
and, discuss its qualitative behavior. A comparison of the different
contributions follows. In the last section we compare experimental results
with these trends and propose new experiments that are able to decide
whether one of these mechanism dominates in real trilayer systems.  
\section{spacer and interface contribution}
The interlayer coupling
energy $J_{\rm inter}$ is usually defined as the difference of grand
canonical potential densities of the parallel and antiparallel aligned system
\cite{Bru95,foot00}:
\begin{equation}
-2J_{\rm inter}=\Omega_{\uparrow\uparrow}-\Omega_{\uparrow\downarrow}\quad.
\end{equation}
To consider the temperature dependence 
one wants to describe the system at a given particle number rather than
at a fixed chemical potential. Therefore the grand canonical potentials have
to be replaced by the free energy densities.
\begin{equation}
-2J_{\rm inter}=F_{\uparrow\uparrow}-F_{\uparrow\downarrow}\quad.
\label{Fdiff}
\end{equation}
Within the quantum well picture it is assumed that the system is a Fermi
liquid, which is correct for the spacer only. 
Furthermore it is assumed that the single particle energies are
temperature independent.
Actually, this is the assumption that excludes the effects of
thermally excited spin waves in this model. Furthermore this assumption
leads to temperature independent reflection coefficients and is justified
 only at temperatures well below the Curie
temperature. 
Finally the norm of the reflection coefficients should vary only
slightly with energy while its argument has to be a continuous function
of energy at the Fermi edge.\\
Now, the crucial quantities for the temperature
dependence are the following\cite{LeC00}:
\begin{itemize}
\item the spacer thickness $D$, or equivalently, the number of spacer
  monolayers $N$, 
\item the stationary Fermi surface spanning vectors parallel to the
  film normal $q_F^\alpha$. Here the index $\alpha$ counts these vectors.
\item the Fermi velocity at these vectors
\begin{eqnarray}
\left(\hbar \nu_{F(+;-)}^\alpha\right)^{-1}=\frac{d k_{z(+;-)}^\alpha}{d
  \epsilon}\Big|_{\epsilon=\epsilon_F}\quad,\nonumber
\end{eqnarray} 
where $k_{z(+;-)}$
denotes the z components of the starting and end point of the spanning
vector, 
\item and the energy derivative of the argument of the 
  reflection coefficient asymmetry $\Delta r^\alpha =|\Delta
  r^\alpha|e^{i\phi^\alpha}$ at the stationary points $\vec{k^\alpha}$
\begin{equation}
D_\phi^\alpha=\frac{ d \phi^\alpha}{d \epsilon}\Big|_{\epsilon=\epsilon_F}\quad .
\label{Deps}
\end{equation}
\end{itemize}
With the restrictions mentioned above the coupling can be written as\cite{LeC00}
\begin{eqnarray}
J_{\rm inter}&=&\sum_\alpha J_{\rm inter}^\alpha(N,0)\cdot f^\alpha(N,T)\quad,
\label{fnt}
\end{eqnarray}
with the temperature dependent functions
\begin{eqnarray}
f^\alpha(N,T)&=&\frac{c_\alpha T}{\sinh(c_\alpha T)}\nonumber\\
c_\alpha&=&a^\alpha N+b^\alpha\quad.
\label{fitbruno}
\end{eqnarray}
Here
\begin{eqnarray}
a^\alpha N&=&\frac{2\pi k_B D}{\hbar \nu_{F}^\alpha}\quad\mbox{depends solely on
  the spacer and}\nonumber\\
b^\alpha&=&2\pi k_B D_\phi^\alpha\quad\mbox{is the interface contribution}
\label{aandb}
\end{eqnarray}
\begin{figure}
\epsfig{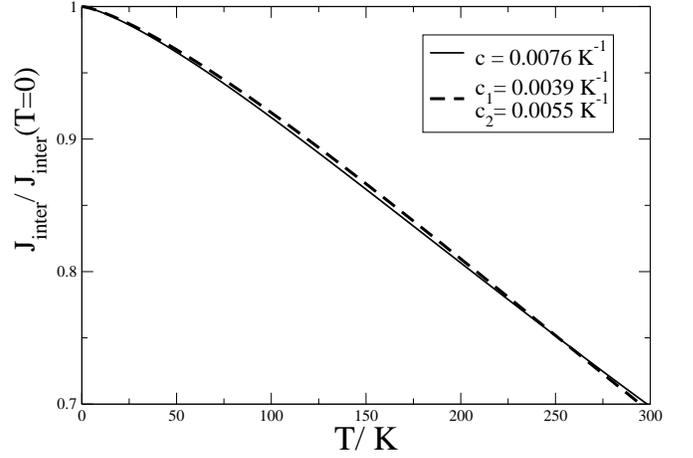}
\caption{The spacer contribution to the temperature dependence of $J_{\rm
    inter}$ according to Eq.(\ref{fnt}) for the case of a Cu(001) spacer with 20 monolayers (dashed
  line). The parameters $\rm c_1,\, c_2$ were taken from Ref. 8, all
  other parameters from Ref. 2. The solid line
  is the function $cT/\sinh(c T)$.}
\label{JIfirst}
\end{figure}
Recall that $\alpha$ counts the number of stationary Fermi surface
spanning vectors and hence the number of oscillation periods in $J_{\rm inter}(N)$\cite{Bru95}.
The spacer contribution constants $a^\alpha$ depend only on the well
known variables $\nu_F^\alpha$ and $d_{\rm sp}=\frac{D}{N}$. They are very
small with a typical order of magnitude of
$a^\alpha\approx 10^{-4}\,K^{-1}$. Ab-initio studies show that the values for $b^\alpha$
are not
considerably higher\cite{DKB99}. Thus $c^\alpha\cdot T$ is a very small
quantity, too, in the temperature regime of interest. We can therefore expand:
\begin{eqnarray}
f^\alpha(c_\alpha\cdot T)&\approx& \frac{1} {1+\frac{1}{6}\left(c_\alpha\cdot
    T\right)^2}\nonumber\\
&\approx& 1-\frac{1}{6}\left(c_\alpha\cdot T\right)^2\left(
  1-\frac{1}{6}\left(c_\alpha\cdot T\right)^2\right)
\label{fvonT}
\end{eqnarray}
This behavior resembles a potential law. The effective exponent $y_\alpha$,
defined as the best fit parameter in
\begin{equation}
f^\alpha(T)\approx 1-x_\alpha\,T^{y_\alpha}\quad,
\label{fit}
\end{equation}
is between one and two ($1<y_\alpha<2$). One can read off from Eq.(\ref{aandb}) that the main difference between the spacer and the
interface contribution is their dependence on the spacer thickness
$D$. While the spacer contribution scales linearly with $D$ the interface
contribution is independent of $D$.\\
Let us discuss the ratio $J_{\rm inter}(T)/J_{\rm inter}(0)$. For the case of
a single oscillation period it is simply given by $f(T)$ from
Eq.(\ref{fitbruno}) or Eq.(\ref{fit}). This simple relation does not
longer hold for more than one oscillation periods. However, as seen in Fig.\ref{JIfirst}, the spacer and interface
  contribution to the temperature dependence is still approximately given by
\begin{equation}
\frac{J_{\rm inter}(T)}{J_{\rm inter}(0)}=\frac{c T}{\sinh(cT)}
\label{approx}
\end{equation}
 and the fit parameter $c$ has the same order of
  magnitude as the parameters $c_\alpha$ from Eq.(\ref{fitbruno}).
\\   
 In the next section we derive the
respective expressions for the magnetic contribution and compare them
with the results described above.
\section{Contribution of magnetic layers}
\begin{figure}
\epsfig{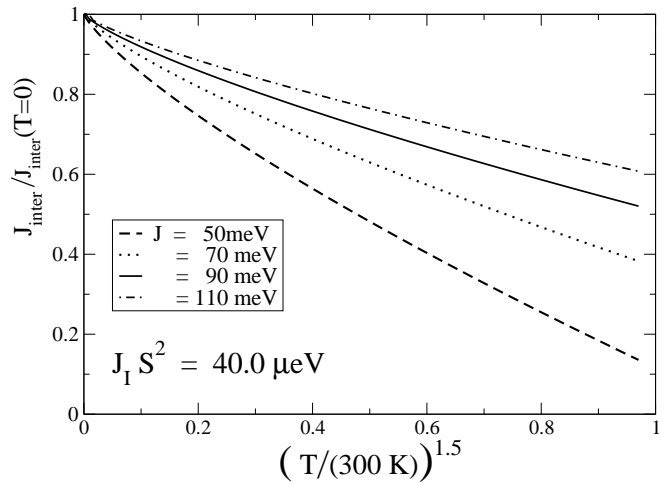}
\caption{Temperature dependent factor of $J_{\rm inter}$ plotted against
  temperature for different intra-layer couplings $J$}  
\label{JI5}
\end{figure}

\begin{figure}[t]
\epsfig{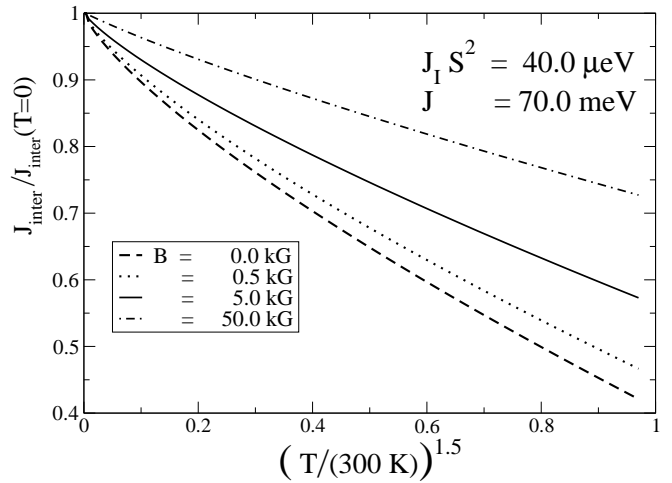}
\caption{Temperature dependent factor of $J_{\rm inter}$ plotted against
  temperature for different external fields.}  
\label{JI8}
\end{figure}

\begin{figure}[t]
\epsfig{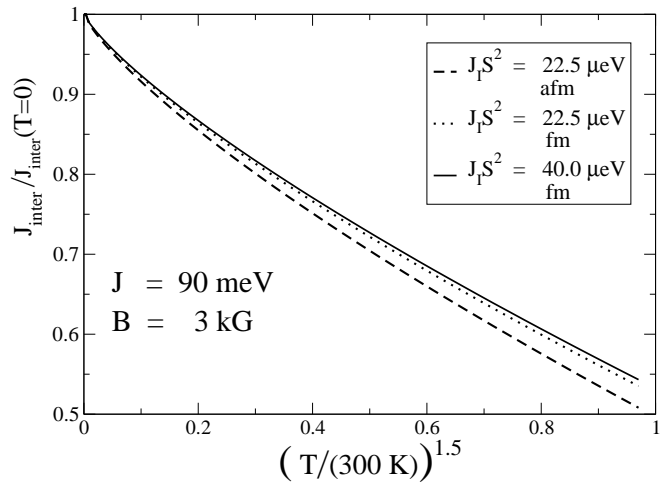}
\caption{Temperature dependent factor of $J_{\rm inter}$ plotted against
  temperature for two ferromagnetic couplings and one anti-ferromagnetic
  coupling.}  
\label{JI7}
\end{figure}

{\bf The model}\\
Our model consists of two equivalent magnetic monolayers A, B with a
ferromagnetic nearest neighbor Heisenberg exchange
\begin{equation}
H_1=-J\sum_{\langle ij\rangle}\left( \vec{S}_{ia}\cdot \vec{S}_{ja}+
  \vec{S}_{ib}\cdot \vec{S}_{jb}\right)\quad J>0\quad.
\end{equation}
The sum runs over all pairs of nearest neighbors within a layer. The
layers are coupled by an interlayer exchange term
\begin{equation}
H_2=-J_I\sum_i \vec{S}_{ia}\cdot \vec{S}_{ib}
\end{equation}
and a magnetic field is added
\begin{equation}
H_3=-{B}^\prime \sum_i \left(S_{iaz}+S_{ibz}\right)\quad.
\end{equation}
${B}^\prime$ is shorthand for $g\mu_B {B}$. The field is strong enough to
align the magnetic moments of both layers parallel, even if the
interlayer coupling $J_I$ is anti-ferromagnetic. This suits the
experimental situation of a ferromagnetic resonance (FMR) experiment in
the saturated limit\cite{LiB03}. The second term describes the interlayer coupling
mediated by the spacer ($J_I>(<)0$ gives (anti)ferromagnetic coupling).
The microscopic constant $J_I$ should be distinguished from the
interlayer coupling energy $J_{\rm inter}$ which is a contribution to
the free energy density of the system as defined in Eq.(\ref{Fdiff}). At
zero temperature $J_I$ and $J_{\rm inter}$ are closely connected and one
finds after a simple and straightforward calculation
\begin{eqnarray}
J_{\rm inter}=J_IS^2\quad.
\label{Teqzero}
\end{eqnarray}
$S$ denotes the spin quantum number.
To account for the temperature dependence resulting from the spacer and
the interfaces one has to replace the constant $J_I$ by an effective,
temperature dependent quantity $J_I\cdot f(N,T)$. 
However, we want to calculate the effect of the magnetic contribution alone and assume
in the following that the mechanisms (i) and (ii)
are unimportant for the considered temperatures. The constant
$J_I$ comprises all important spacer and interface properties at zero
temperature as
e.g. spacer thickness, spacer material, geometry, interface roughness
and so on.
The whole Hamiltonian is the sum of all terms above
\begin{eqnarray}
H=H_1+H_2+H_3\quad.
\label{Ham}
\end{eqnarray}
The same model was studied by Almeida, Mills, and Teitelmann\cite{AMT95} to get
information about the interlayer exchange coupling. However, they
discuss the temperature dependence of the spin wave excitations within
a renormalized spin wave theory following Dyson\cite{Dys56}. In this theory the
spin wave excitations can be described by effective, temperature
dependent coupling "constants" $J_I^\star(T)$ and $J^\star(T)$. In
Ref. 12 the temperature dependence of $J_I^\star$ is
discussed.\\
But note that in our case the crucial quantity is not $J_I^\star(T)$ but the interlayer coupling energy $J_{\rm
  inter}(T)$ as defined in Eq.(\ref{Fdiff}). One has to distinguish
carefully between both variables. An important difference is that the
temperature variation of $J_I^\star(T)$ is caused by {\it interactions} of spin
waves, while the mere {\it excitation} of spin waves already reduces  $J_{\rm
  inter}(T)$.\\ We will now describe how $J_{\rm inter}(T)$ is extracted
from our model and present analytical as well as numerical results.\\\\
{\bf The coupling}\\
We solve the Hamiltonian (\ref{Ham}) within the free spin wave
approximation which is a good treatment for low temperatures and is
correct for zero temperature. Using the Holstein-Primakoff transformation\cite{HoP40}
we obtain a bosonic Hamiltonian that describes spin waves in the magnetic
sheets A and B:
\begin{eqnarray}
H&=&
E_0+\sum_\vec{q}\left[D_{1\vec{q}}\left(n_{\vec{q}a}+n_{\vec{q}b}\right)+D_2\left(a_\vec{q}^\dagger
    b_\vec{q}+b_\vec{q}^\dagger a_\vec{q}\right)\right]\nonumber
\end{eqnarray}
\begin{eqnarray}
\frac{1}{N}E_0&=&-J_IS^2-2JpS^2-2B^\prime S\nonumber\\
D_{1\vec{q}}&=&2J_\vec{q}S+J_IS+B^\prime\nonumber\\
D_2&=&-J_IS\quad.
\label{Hamsw}
\end{eqnarray}
$a(b)_\vec{q}^\dagger$ creates a spin wave with wave number $\vec{q}$ in the layer
$A(B)$. $n_\vec{q}^{a(b)}$ is the respective spin wave density. $p$
denotes the in-plane coordination number. $J_\vec{q}$ is an
abbreviation for $J(p-\gamma_\vec{q})$, and $\gamma_\vec{q}$ is a geometrical factor,
\begin{eqnarray}
\gamma_\vec{q}=\sum_{\vec{\Delta}}e^{i\vec{q}\vec{\Delta}}\quad,\nonumber
\end{eqnarray}
with $\vec{\Delta}$ denoting a vector between nearest neighbors within a
layer. The
new Hamiltonian is bilinear and can be solved exactly, for instance by a
Bogoliubov transformation.  Thus one obtains the single particle excitation
energies
\begin{eqnarray}
 \omega_{\vec{q}_+}&=& 2 J_{\vec{q}}S+B^\prime\nonumber\\
\omega_{\vec{q}_-}&=& 2 J_{\vec{q}}S+B^\prime+2J_IS\quad,
\end{eqnarray}
and the ground state energy $E_0$ from Eq.(\ref{Hamsw}). 
For anti-ferromagnetic coupling a minimal field 
$B^\prime=|2J_IS|$ is needed to avoid negative excitation energies.
To define the interlayer exchange coupling we follow, e.g., Ref. 15
where $J_{\rm inter}$ is treated as a contribution to the free energy density
\begin{eqnarray}
F&=&F_0+F_{\rm ex}\nonumber\\
F_{\rm ex}&=&-J_{\rm inter}\cos(\phi)\quad
\phi=\angle(\vec{M}_A,\vec{M}_B)
\label{Def}
\end{eqnarray}
Inserting $\phi_1=0$ and $\phi_2=\pi$ into this expression we immediately
arrive at the definition (\ref{Fdiff}) used in the quantum well picture and in
ab-initio theory. However, for finite coupling ($J_I\neq 0$) one of these
angles is not the equilibrium angle. The respective
configuration is unstable against spin wave excitation, which may cause
problems in the evaluation of Eq.(\ref{Fdiff}). To avoid these
complications we evaluate Eq.(\ref{Def}) directly. $F_0$ is the part of the free energy density that is not
connected with the interlayer coupling. It can be obtained immediately
using
\begin{equation}
F_0=F(J_{\rm I}=0)\quad.
\label{F0}
\end{equation}
Here $F(J_I=0)$ is the free energy density of the uncoupled system where the coupling $J_{\rm I}$ is set to zero while all the other
parameters are the same as in the full system. Since we consider a
parallel alignment of all magnetic moments in the ground state
($\phi=0$) we simply get
\begin{equation}
-J_{\rm inter}=F_{\rm ex}= F-F(J_{\rm I}=0)
\label{Jinterdef}
\end{equation}
For the free energy densities of the full and the uncoupled system we
find, respectively,
\begin{eqnarray}
L\cdot F&=&-k_BT\ln{\Xi}=E_0+k_BT\sum_\vec{q}\big[\ln\left(1-e^{-\beta\omega_{\vec{q}_+}}\right)+\nonumber\\
&{}&\qquad\qquad\qquad\qquad\qquad\qquad+\ln\left(1-e^{-\beta\omega{\vec{q}_-}}\right)\big]\nonumber\\
L\cdot F_0&=&-k_BT\ln\Xi_0=E_0(J_I=0)+\nonumber\\
&{}&\quad\qquad\qquad\qquad+k_BT\sum_\vec{q}2\ln\left(1-e^{-\beta\omega_{\vec{q}_+}}\right)\nonumber
\end{eqnarray}
$L$ is the size of the system, i.e. the number of sites within a
layer. $\Xi_{(0)}$ denotes the partition function. Note that in our
model the chemical potential is equal to zero.  Consequently the free
energy is identical to the grand canonical potential which justifies the
equations above. 
The interlayer exchange coupling finally reads
\begin{eqnarray}
J_{\rm inter}&=&
J_IS^2-k_BT\frac{1}{L}\sum_\vec{q}\left|\ln\left(1-e^{-\beta\omega_{\vec{q}_+}}\right)\right|\nonumber\\
&{}&\qquad\qquad\qquad-\left|\ln\left(1-e^{-\beta\omega{\vec{q}_-}}\right)\right|\quad. 
\label{Jinter}
\end{eqnarray}
\begin{figure}
\epsfig{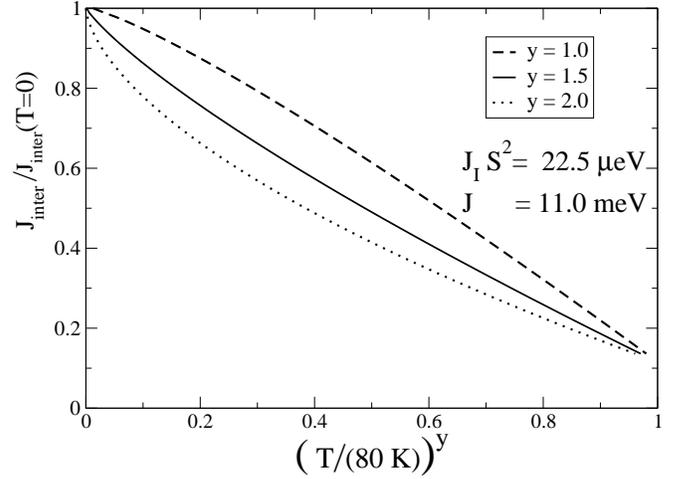}
\caption{Temperature dependent factor of $J_{\rm inter}$ plotted against
  temperature in different scales. Here and in the following figures
  solely the {\it magnetic} contribution to the temperature dependence is shown.}
\label{JI1}
\end{figure} 
This equation can be easily evaluated.  
Furthermore an analytical expression can be derived:\\ Let us assume, e.g., a quadratic lattice. The dominating terms in the sum over the
two dimensional Brillouin zone stem from the vicinity of the
$\Gamma$-point where $q$ is small and we can write: $ J_{\vec{q}}\approx
Jq^2$, where $q$ is the norm of $\vec{q}$. After expanding the logarithm and replacing the $\vec{q}$-summation by
an integral we get
\begin{eqnarray}
J_{\rm inter}&\approx&J_IS^2-k_BT\sum_{n=1}^\infty\frac{1}{n}e^{-\beta B^\prime
  n}\left(1-e^{-\beta\cdot 2 J_I
    Sn}\right)\cdot\nonumber\\
&{}&\cdot \frac{1}{2\pi}\int_0^{q_0}\,dq\,qe^{-\beta\,2JSnq^2}
\end{eqnarray}
The integral is written in polar coordinates $(q,\phi)$ and the
trivial $\phi$ integration has already been performed. $q_0$ is the averaged
extension of the first Brillouin zone. Since terms with large values of
$q$ only contribute negligibly to the integral, we may approximately
replace the upper limit by infinity and use the tabulated integral
$\int_0^\infty\,dt\,te^{-\alpha t}=(2\alpha)^{-1}$. Thus we end up with
\begin{eqnarray}
f(T)&=&\frac{J_{\rm inter}(T)}{J_{\rm inter}(0)}\nonumber\\
&=&
1-\frac{1}{8\pi
  JS}\frac{1}{J_{\rm inter}(0)}\left(k_B T\right)^2\cdot
\Sigma(T)\quad,\nonumber\\
\Sigma(T)&=&\sum_{n=1}^\infty\frac{1}{n^2}e^{-\beta
  B n}\left(1-e^{-\frac{1}{S}J_{\rm inter}(0)\beta n}\right)\quad.
\label{Jinter2}
\end{eqnarray}
The infinite sum converges by the majorant criterion (note the constraint $B^\prime>|2J_IS|$ for anti-ferromagnetic
coupling).
The first derivative of $\Sigma(T)$ with respect to $T$ is negative\cite{foot1},
while the first derivative of the term $k_B T\cdot \Sigma(T)$ is larger
than zero. 
Thus the coupling decreases with temperature faster than $1-x\; T$ but
slower than $1-x\; T^2$. The effective coefficient $y$, defined in
Eq. (\ref{fit}), is between 1 and 2.
The evaluation of Eq.(\ref{Jinter}) clearly
corroborates this trend as can be seen in Fig.\ref{JI1}. The effective coefficient is around $y=1.5$ except
for very low temperatures $T<30\,K$. 
Here and in the following calculations the parameters $J_I$
are chosen to be comparable with experiment
\cite{LiB03,LiB03a} using Eq.(\ref{Teqzero}). The effective intra-layer coupling $J$ is chosen such
that the spin wave stiffness of the bulk material has a realistic order of magnitude
($J=10-100~{\rm meV}$ for transition metals\cite{PKT01}).
For this parameters we find a certain decrease of $J_{\rm inter}$ between $0$ and $300\,K$.
\\
\begin{figure}
\epsfig{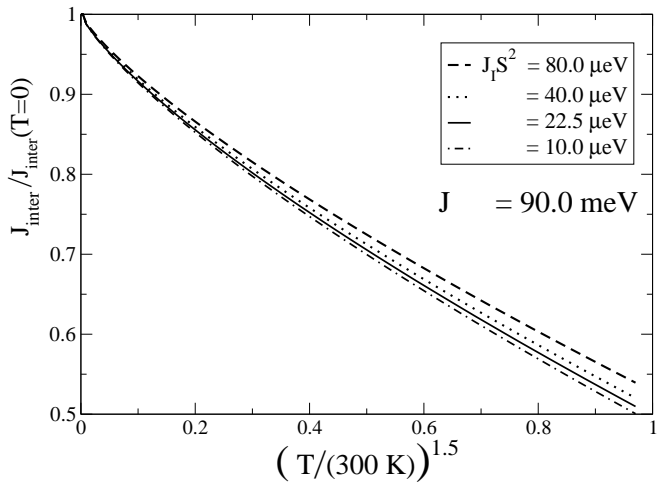}
\caption{Temperature dependent factor of $J_{\rm inter}$ plotted against
  temperature for different zero temperature couplings $J_IS^2$.}  
\label{JI4}
\end{figure}
Fig.\ref{JI4} shows the dependence of $f(T)$ on the zero temperature
coupling $J_{\rm inter}(0)$. The temperature dependence is more
pronounced if $J_{\rm inter}(0)$ is small. However, the differences
between the curves are
very small. $J_{\rm inter}(0)$ appears twice in Eq.(\ref{Jinter2}), once in the denominator and once in the exponent. These
contributions seem to cancel each other almost perfectly.\\
The dependence on the {\it intra}layer coupling $J$ is much more
pronounced. This is seen in Eq.(\ref{Jinter2}) as well as in
Fig.\ref{JI5}. Materials with a large effective coupling $J$ have a much less pronounced temperature
dependence.\\ In addition the function $f(T)$ depends
on the external field $B$ (Fig.\ref{JI8}). External fields stabilize
the coupling, since more energy is needed to excite a
magnon and the ground state is stabilized.\\
This property also
influences the dependence of $f(T)$ on the coupling sign. One
can read off from Eq.(\ref{Jinter2}):
\begin{eqnarray}
&f&\big(T,{\rm fm},B^\prime-2J_IS\big)\nonumber\\
=&f&\big(T,{\rm afm},B^\prime\big)\quad,
\end{eqnarray}
which means, that for anti-ferromagnetic coupling an effective field 
\begin{equation}
B_{\rm eff}=B^\prime-2J_IS\quad,
\end{equation}
rather than the pure external field, is decisive. Thus the temperature
dependence is more pronounced for anti-ferromagnetic coupling compared
with ferromagnetic coupling. This is shown in Fig.\ref{JI7}, where
results for anti-ferromagnetic and ferromagnetic coupling are shown for
$|J_{\rm inter}(0)|=22.5\,\mu eV$. For comparison a curve for $J_{\rm
  inter}(0)=+40\,\mu eV$ is shown. The dependence on the norm of $J_{\rm
  inter}$ is almost negligible compared to the influence of the
sign. However, this very influence is rather weak, too.\\
The behavior of the magnetic contribution worked out above, will be
compared to the spacer and
interface contribution in the next section.
\section{Comparison of the different coupling mechanisms}
The spacer, interface and magnetic contribution show some similarities:
\begin{itemize}
\item In the temperature regime where the theories are applicable the functional dependencies $J_{\rm
    inter}(T)$ resemble each other. We have for all contributions
\begin{equation}
f(T)=\frac{J_{\rm inter}(T)}{J_{\rm inter}(0)}\approx 1-x\,T^y\quad,
\end{equation}
with $1<y<2$.
\end{itemize}
There are, however, certain differences:
\begin{itemize}
\item The dependence of $f(T)$ on the spacer thickness $D$ is quite
  different. The spacer mechanism exhibits a strict $D\cdot T$
  dependence
\begin{equation}
f^{\rm spacer}(D,T)=f(D\cdot T)\quad,
\end{equation}
the interface contribution is independent of $D$,
\begin{equation}
f^{\rm interface}(D,T)=f(T)\quad,
\end{equation}
while the magnetic layer contribution shows a very weak  implicit
dependence via the zero temperature coupling
\begin{equation}
f^{\rm magnet}(D,T)=f(J_{\rm inter}(0,D),T)\quad.
\end{equation}
that oscillates with the spacer thickness.
\item
There are also differences concerning the dependence on the magnetic material. The spacer
contribution is independent of the
magnetic material, the interface contribution may be material dependent
via $D_\phi$ and the magnetic contribution exhibits a strong $\frac{1}{J}$
dependence, where $J$ is the effective coupling between the magnetic
moments of the film.
\item
The magnetic contribution shows a (weak) dependence on the coupling
sign, i.e. the temperature dependence is more pronounced for
anti-ferromagnetic interlayer coupling, if the coupling strength is the same.
\item
The magnetic contribution is suppressed by an external field. To our
knowledge, there is no such effect for the spacer or interface
contribution.
\item 
Alloying of the spacer introduces disorder and can reduce the amplitude
of the coupling\cite{DKB98}. In this case the temperature dependence of
the spacer contribution is {\it reduced} (see Ref. 18), while the
temperature dependence of the magnetic contribution is {\it increased} (see
Fig.\ref{JI4}). However, this statement has to be taken with care, since
alloying may also change the stationary Fermi surface spanning vectors
and therewith the parameters $c_\alpha$ of Eq.(\ref{fitbruno}). For this
case alloying may be increase or decrease the spacer contribution depending on
the specific combination of materials.

\end{itemize}
The specific behavior of the different mechanisms opens the
possibility to identify the dominant mechanism by experiments. To this
end we will
review existing experiments and
propose new experiments in the next section.
\section{Experiments}
\begin{figure}[t]
\epsfig{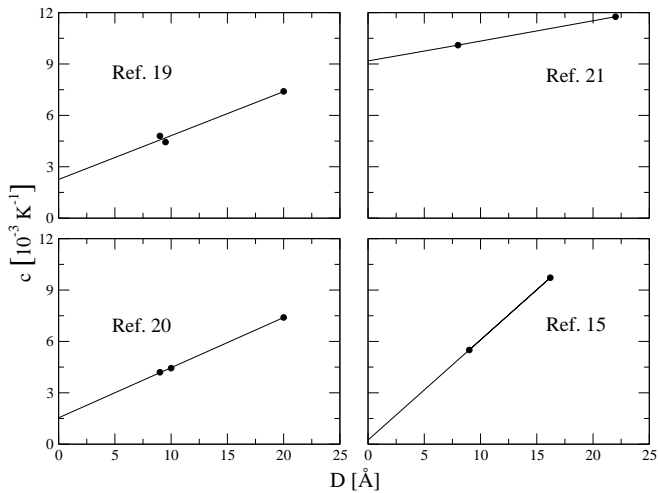}
\caption{Parameter $c$ displayed against spacer thickness as obtained by
  different experiments. Large values of $c$
  mean pronounced temperature dependence. The solid lines are linear
  extrapolations to $D=0$}  
\label{JI00}
\end{figure}

There are not many reported studies dealing with the temperature
dependence $f(T)$ of the interlayer coupling. Z.~Zhang et.al. \cite{ZZW94a,ZZW94}
studied ${\rm Co_{\rm 32\,\AA}/Ru_{\rm x\,\AA}/Co_{\rm 32\, \AA}}$ trilayers using
ferromagnetic resonance. N.~Perat and A.~Dina studied ${\rm Co_{\rm
    24\AA}/Cu_{x\AA}/Co_{\rm 24\AA}(hcp)}$
samples using squid magnetization measurements\cite{PeD97}. J.~Lindner and
K.~Baberschke performed ferromagnetic resonance measurements on a
${\rm Ni_7/Cu_x/Co_2(001)}$ system\cite{LiB03a}.\\ With one exception (Ref. 19,20: $D=24\,\AA$) all data can be fitted
to Eq.(\ref{fitbruno}). 
In all cases the parameter $c$
deviates clearly from the value expected from the spacer contribution
(\ref{aandb}) alone. It was further shown by Lindner et.al.\cite{LRK02} that
the data can be fitted with the same accuracy to Eq.(\ref{fit})
with $y=1.5$. Both functional behaviors fulfill our expectation
and can be caused by any of the described mechanisms. As discussed in
detail above we have to know the dependence on the intra-layer coupling $J$ or
on the spacer thickness to discriminate between the different
contributions. Unfortunately the dependence on the magnetic material
(and therewith on $J$) was
not investigated in these studies. On the other hand, there are some
data describing the influence of the spacer thickness. They are summarized in
Fig.\ref{JI00}.
 There are two parameters that are a measure of the thickness
dependence of $f(T)$, namely the parameter $x$ from
Eq.(\ref{fit}) and the parameter $c$ from Eq.(\ref{fitbruno}). Large $x$ or large $c$ indicate a large
suppression of the coupling by temperature. In Fig.\ref{JI00} the
parameter $c$ is
displayed rather than $x$ since it is more convenient to obtain its values
from the experimental studies. The data points were taken directly
from the papers or were extracted from the respective plots.\\ The parameter
$c$ increases with the spacer thickness $D$ in all cases. This qualitative trend is in
accordance with the spacer but also with the magnetic contribution. A
linear increase would favor a strong importance of the spacer
contribution, while oscillations that follow $J_{\rm inter}(0,D)$ would
indicate a decisive role of the magnetic mechanism. However, in all works there are not
enough data points to establish a linear or oscillatory behavior.\\
If one assumes for a moment a linear dependence according to
Eq. (\ref{aandb}) the solid lines in Fig.\ref{JI00} are obtained. The
graphs of Refs. 19, 20 and 21 show a
certain finite value for the $D=0$ extrapolation. Thus the spacer
mechanism can not be the only source of temperature dependence in these samples.
The spacer thickness dependence is very weak in Ref. 21 as
expected by the magnetic contribution (indeed $J_{\rm inter}(0)$ is very
similar for both data points). On the other hand the value $a$
from Eq.(\ref{aandb}), which can be read off from the slope in
Fig.\ref{JI00}, is in rather good agreement with model theory:
\begin{equation}
a_{\rm ex}\approx2.4\cdot 10^{-4}\,K^{-1}\quad\quad a_{\rm th}\approx
1\cdot 10^{-4}\,K^{-1}
\end{equation}
The theoretical value is taken from Ref. 2.\\
The situation in the ruthenium samples\cite{ZZW94a,ZZW94} seems to be
different. The contribution scaling with the spacer thickness is more important. There is a very interesting
feature in the upper left panel of Fig.\ref{JI00}. There seems to be evidence for a slight
oscillatory behavior of $c$ as a function of spacer thickness. The
oscillation follows the $J_{\rm inter}(0)$ value. 
For the spacer thicknesses of Ref. 21 no oscillations of $J_{\rm
  inter}(0)$with spacer thickness $D$ are found and consequently no
oscillation of $c$. This behavior favors a magnetic mechanism. On
the other hand the fitted $a$ value from Eq.(\ref{aandb}) is again
in reasonable agreement with the theoretical result\footnote{to estimate $c_{\rm th}$ the average Fermi velocity of
  ruthenium $\approx 6\cdot10^7 {\rm ~cm\,s^{-1}}$ was taken.}

\begin{equation}
a_{\rm ex}\approx5\cdot 10^{-4}\,K^{-1}\quad\quad a_{\rm th}\approx
2.4\cdot 10^{-4}\,K^{-1}\quad.
\end{equation}
The deviations of a factor 2-3 are not alarming, since the linear fits
are of course of bad quality due to the small number of data points.\\
The data of Ref. 15 reveal a different picture. Here the
parameter $c$ really seems to scale with the
spacer thickness as predicted by the model theory of the spacer
contribution. Of course two points are not enough to confirm this
mechanism and the value of $a$ differs from the theoretical one by an order
of magnitude.
\begin{equation}
a_{\rm ex}\approx2.4\cdot 10^{-3}\,K^{-1}\quad\quad a_{\rm th}\approx
1\cdot 10^{-4}\,K^{-1}\quad.
\end{equation}
Again the theoretical value is taken from Ref. 2.
This system was also investigated by ab-initio calculations \cite{DKB99} corroborating
the order of magnitude of $a_{\rm th}$. Thus the origin of the strong
difference remains unclear.\\
In summary, no clear conclusion can be drawn from the existing
experiments. There is clearly a need for more experimental data. We
propose a systematic investigation of the temperature dependence at
different spacer thicknesses. The spacer thickness should be varied at
least over a full oscillation of $J_{\rm inter}$ with $D$. The
parameters $c$ of Eq.(\ref{fitbruno}) or $x$ of Eq.(\ref{fit})
should be displayed as a function of spacer thickness $D$ and as a function
of the zero temperature coupling $J_{\rm inter}(0)$.\\
In addition we propose the study of the temperature dependence for
different magnetic materials (e.g. Co, Ni) separated by the same spacer
(e.g. Cu).\\
With these experimental results at hand and with the theoretical results summarized in section 4
one may isolate the dominating mechanism that causes the temperature
dependence of the interlayer coupling in metallic trilayers.\\
There is clearly a need for more theoretical studies as well. Both
aspects, the spacer and interface contribution on the one hand and the
magnetic contribution on the other, should be described in one model on
equal footing. Furthermore the restriction to low temperatures, which is
up to
now inherent to all models, should be removed and effects as the
temperature dependence of the reflection coefficients should be studied
as well.

\section{Summary}
The reduction of the interlayer coupling with temperature in metallic
multilayers may be caused by effects within the spacer, at the interface,
or within the magnetic layers. We derived the magnetic part at low
temperatures and discussed its dependence on the spacer layer thickness,
on the magnetic materials, on the sign of the coupling, and on
the external field. These dependencies were compared with those of the spacer and interface contributions. As a main result we found
that the functional dependence of the temperature dependent factor $f(T)$
is roughly the same for all mechanisms. There are certain differences in
the dependence of $f(T)$ on the spacer thickness and on the magnetic
material. Based on these considerations we proposed experiments that are
able to identify the dominant mechanism in metallic trilayers which is
not possible with the experimental data available today.

\section*{Acknowledgments}
This work is supported by the Deutsche Forschungsgemeinschaft within
the Sonderforschungsbereich 290.
Fruitful discussions with K.~Baberschke are gratefully acknowledged.


\begin{thebibliography}{10}
\bibitem{foot00}
There is no unique conventien concerning the sign and the prefactor in
this definition.
\bibitem{foot1}
this holds only for $B<k_BT$ which is usually fulfilled in
experiments for $T>1K$
\end{thebibliography}
\end{document}